\documentclass{article}
\usepackage{spconf}
\usepackage{amsmath,graphicx}
\usepackage{amssymb}
\usepackage{amsfonts}
\usepackage{algorithm}
\usepackage{amsthm}
\usepackage{cite}
\usepackage{url}
\usepackage{epsfig}
\usepackage{subfigure}
\usepackage{psfrag}
\usepackage{color}

\title{Joint Data Association, NLOS Mitigation, and Clutter Suppression for Networked Device-Free Sensing in 6G Cellular Network}
\name{Qin Shi, Liang Liu, Shuowen Zhang}
\address{EIE Department, The Hong Kong Polytechnic University\\
	Email: qin-eie.shi@connect.polyu.hk, liang-eie.liu@polyu.edu.hk, shuowen.zhang@polyu.edu.hk }
\begin{document}
\ninept
\maketitle
\begin{abstract}
Recently, there is a growing interest in achieving integrated sensing and communication (ISAC) in the sixth-generation (6G) cellular network. Inspired by this trend and the success of cooperative communication in cloud radio access network, this paper considers a networked device-free sensing architecture based on base station (BS) cooperation to transform the cellular network into a huge sensor that can provide ubiquitous and high-performance sensing services. Under this framework, the BSs first transmit the downlink communication signals to the mobile users and then estimate the range information of the targets based on their echoes. Next, a central processor collects the range information from all the BSs via the fronthaul links and localizes each target based on its distances to various BSs. To enable the above strategy in the 6G network, we will perform joint data association, non-line-of-sight (NLOS) mitigation, and clutter suppression, such that the central processor is able to find out the useful range estimations extracted from the line-of-sight (LOS) paths and match them to the right targets for localization. Numerical results show that our interested networked device-free sensing scheme for the 6G network can localize the targets with high accuracy in the challenging multi-path propagation environment.
\end{abstract}

\begin{keywords}
Integrated sensing and communication (ISAC), networked sensing, 6G, data association, non-line-of-sight (NLOS) mitigation, clutter suppression.
\end{keywords}
\newtheorem{example}{Example}
\newtheorem{corollary}{Corollary}
\newtheorem{definition}{Definition}
\newtheorem{lemma}{Lemma}
\newtheorem{theorem}{Theorem}
\newtheorem{proposition}{Proposition}
\newtheorem{remark}{Remark}
\newcommand{\mv}[1]{\mbox{\boldmath{$ #1 $}}}

\section{Introduction}\label{sec:Introduction}
Thanks to the wide bandwidth at the millimeter wave/Terahertz band and the large antenna array brought by the massive multiple-input multiple-output (MIMO) technique, the cellular network tends to be capable of sensing the environment via the communication signals with ultra-high range and angle resolution, similar to the radar systems. Hence, there is a recent trend in both the academia and the industry to achieve integrated sensing and communication (ISAC) in the sixth-generation (6G) cellular network, where the base stations (BSs) can employ a common wireless signal for conveying information to the mobile users and localizing the targets simultaneously \cite{Liu22,Zhang21,Tan21,Vorobyov19,Hassanien19,paul2016survey,zheng2019radar,liu2020joint,An22}. It is anticipated that the 6G-enabled ISAC technique will play an important role in an enormous number of newly emerging applications from smart transportation systems, smart factories, etc., where both the sensing function and the communication function are of paramount significance.

In the literature of the ISAC technology, the performance trade-off between the capacity in communication and the estimation distortion in sensing have been optimized in various works \cite{Masouros18radar,Mu22,Tsinos21,Eldar20}, because the optimal waveforms for the communication signals and the sensing signals are quite different \cite{sturm2011waveform}. Apart from the performance optimization, several interesting works have been done to design the practical signal processing techniques for embedding the sensing function into the 6G cellular network. For example, efficient algorithms have been proposed such that a BS can extract the range/angle/Doppler information of the targets based on the orthogonal frequency division multiplexing (OFDM) signals \cite{Wang17,Liu20MIMO}, the orthogonal frequency time space (OFTS) signals \cite{Gaudio20}, and the millimeter wave signals \cite{Dokhanchi19}, that are reflected by these targets. Moreover, \cite{Barneto22,Yang22} have devised powerful estimation schemes such that a mobile user can utilize the cellular signals for realizing simultaneous localization and mapping (SLAM).

\begin{figure}
\centering
\includegraphics[scale=0.185]{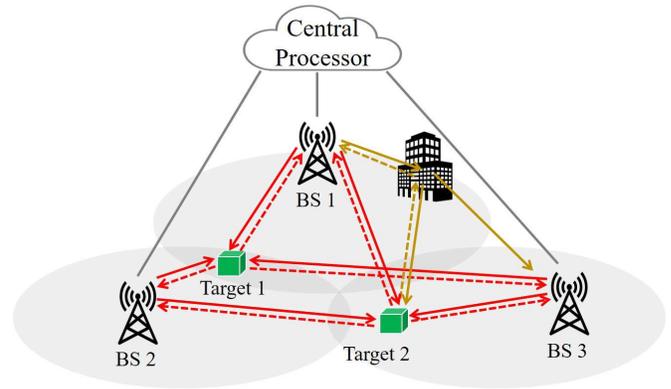}
\caption{System model of our considered networked device-free sensing architecture. The BSs are connected to the central processor via fronthaul links to share the range estimations. For wireless propagation, there are LOS paths (e.g., the path from BS $1$ to Target $1$ to BS $2$), NLOS paths (e.g., the path from BS $1$ to the building to Target $2$ back to BS $1$), and the paths arising from the clutters (e.g., the path from BS $1$ to the building to BS $3$).}\vspace{-10pt}
\label{fig1}
\end{figure}

It is worth noting that the above works mainly consider the scenario where localization is performed with one transmitter and one collocated/separate receiver, as in the monostatic/bistatic radar systems. Inspired by the cloud radio access network where the BSs can collaborate with each other to mitigate the inter-cell interference, our recent work \cite{shi2022device} proposed a novel networked device-free sensing architecture, as shown in Fig. \ref{fig1}. Specifically, all the BSs in a certain area first emit the OFDM signals in the downlink, then estimate the range information of all the targets based on their echoes, and at last send their range estimations to a central processor via the fronthaul links. Next, the central processor can localize each target based on its distances to various BSs. As pointed out by \cite{shi2022device}, one main challenge of the above scheme lies in data association, i.e., how to match the range information in each received echo to the right target. Under the line-of-sight (LOS) propagation environment, an efficient data association algorithm was proposed in \cite{shi2022device} to enable networked device-free sensing.

In this paper, we generalize our results in \cite{shi2022device} to a more practical but also more challenging multi-path propagation environment \cite{Aditya18,Guvenc09}, where besides the LOS paths, non-line-of-sight (NLOS) paths and paths arising from clutters may exist as well among the BSs and the targets, as shown in Fig. \ref{fig1}. Under the two-phase networked device-free sensing protocol, we design an efficient range estimation algorithm based on the OFDM channel estimation technique in Phase I, and perform joint data association, NLOS mitigation, and clutter suppression in Phase II. After the useful range estimations from the LOS paths are identified and each of them is matched to the right target, we can localize these targets accurately. Numerical results verify that our proposed scheme can achieve very high localization accuracy with low complexity.

\section{System Model}\label{sec:model}
In this paper, we consider a 6G-enabled ISAC system consisting of $M$ BSs, $K$ targets to be localized (the value of $K$ is unknown and needs to be estimated), and $I$ users for communication. Since the communication technology is very mature in the cellular network, we mainly focus on the sensing function in this ISAC system. Let $(a_m,b_m)$ and $(x_k,y_k)$ denote the 2D coordinates of the $m$-th BS and the $k$-th target, respectively, $m=1,\ldots,M$, $k=1,\ldots,K$. Then, the distance between the $m$-th BS and the $k$-th target is given by
\begin{align}
d_{m,k}&=f(x_k,y_k,a_m,b_m)\nonumber \\ &=\sqrt{(a_m-x_k)^2+(b_m-y_k)^2},~\forall m, k.
\label{eqS2.1}
\end{align}Moreover, the sum of the distance between the $u$-th BS and the $k$-th target and that between the $k$-th target and the $m$-th BS is
\begin{align}
d_{u,m,k}=d_{u,k}+d_{m,k},~\forall  u, m, k.
\label{eqS2.2}
\end{align}In the downlink, the BSs will transmit the OFDM signals to the information receivers, while these signals can be reflected by the targets to different BSs as well. Based on the propagation delay from BS $u$ to target $k$ to BS $m$, we can estimate $d_{m,k}$ if $u=m$ and $d_{u,m,k}$ if $u\neq m$. Then, each target can be localized based on its estimated ranges to various BSs.

Specifically, let $\boldsymbol{s}_m = [s_{m,1},\ldots,s_{m,N}]^T$ denote one frequency-domain OFDM symbol at the $m$-th BS, $\forall m$, where $s_{m,n}$ is the signal at the $n$-th sub-carrier and $N$ is the number of sub-carriers. Then, the time-domain modulated signal of BS $m$ over one OFDM symbol consisting of $N$ samples is given by $\boldsymbol{\chi}_m = [\chi_{m,1},...,\chi_{m,N}]^T = \sqrt{p}\boldsymbol{W}^H \boldsymbol{s}_m$, $\forall m$, where $p$ denotes the common transmit power at the BSs, and $\boldsymbol{W} \in \mathbb{C}^{N \times N}$ denotes the discrete Fourier transform (DFT) matrix with $\boldsymbol{W}\boldsymbol{W}^H=\boldsymbol{W}^H\boldsymbol{W}=\boldsymbol{I}$. After inserting the cyclic prefix (CP) consisting of $Q$ OFDM samples, the time-domain signal transmitted by BS $m$ over one OFDM symbol period is given by $\bar{\boldsymbol{\chi}}_{m}=[\bar{\chi}_{m,-Q},\ldots,\bar{\chi}_{m,-1},\bar{\chi}_{m,0},\ldots,\bar{\chi}_{m,N-1}]$, where if $n>0$, $\bar{\chi}_{m,n} = \chi_{m,n+1}$ denotes the useful signal, and if $n\leq 0$, $\bar{\chi}_{m,n} = \chi_{m,N+n+1}$ denotes the CP.

Define $\boldsymbol{h}_{u,m}=[h_{u,m,1},\ldots,h_{u,m,L}]^T$ as the $L$-tap multi-path channel from BS $u$ to BS $m$, where $h_{u,m,l}$ denotes the complex channel coefficient of the path with a delay of $l$ OFDM sample periods. Then, the received signal at the $m$-th BS in the $n$-th OFDM sample period can be expressed as
\begin{align}
	y_{m,n}=\sum_{u=1}^M \sum_{l=1}^L h_{u,m,l} \bar{\chi}_{u,n-l}+z_{m,n}, ~\forall m,n,
\label{eqS2.4}
\end{align}where $L$ denotes the maximum number of resolvable paths and $z_{m,n} \sim \mathcal {CN}(0,\sigma_z^2)$ denotes the noise at the $m$-th BS in the $n$-th OFDM sample period. Note that each BS $m$ can potentially receive the signals transmitted by BS $u$ via three types of paths - {\bf Type I path}: the LOS path from BS $u$ to some target to BS $m$; {\bf Type II path}: the NLOS path from BS $u$ to BS $m$ via some target and some other reflector/scatter; {\bf Type III path}: the path from BS $u$ to some clutter to BS $m$. Thereby, in (\ref{eqS2.4}), $h_{u,m,l}\neq 0$ indicates that there exists a Type I/II/III path from BS $u$ to BS $m$, whose propagation delay is of $l$ OFDM sample periods; and $h_{u,m,l}=0$ otherwise. Note that if $h_{u,m,l}\neq 0$ is contributed by a Type I path from BS $u$ to some target $k$ to BS $m$, then $d_{u,m,k}/c_0$ is equal to $l$ OFDM sample periods, where $c_0$ is the speed of the light. Therefore, the range information about the targets can be obtained by estimating the time-domain OFDM channels.

Based on the above observation, in this paper, we adopt a two-phase networked device-free sensing protocol \cite{shi2022device}. Specifically, in Phase I, each BS $m$ estimates the time-domain OFDM channels, obtains the propagation delay information of each of its received signals based on the estimated channels, and send its delay (thus range) information to a central processor. However, the central processor does not know which ranges are obtained from the Type I paths that are useful for localization. Moreover, if a range is estimated from a Type I path, it is a challenging job to match the range of this path to the right target that reflects the signal on this path, as shown in \cite{shi2022device}. In Phase II, the central processor thus needs to perform joint data association, NLOS mitigation, clutter suppression so as to identify the signals from the Type I paths that are useful for estimating $d_{u,m,k}$'s and match the range estimations of $d_{u,m,k}$'s to the right targets. Then, the number of the targets, i.e., $K$, can be estimated, and each of the $K$ targets can be efficiently localized based on its distances to various BSs. In the following two sections, we show the details about Phase I and Phase II under the above protocol.

\section{Phase I: Range Estimation}\label{sec:Phase I}
In this section, we show how to perform range estimation in Phase I of our considered two-phase localization protocol. It can be shown that the frequency-domain signal received at BS $m$ can be expressed as \cite{Hwang09,shi2022device}
\begin{align}
 \bar{\boldsymbol{y}}_m \!=\! \sqrt{p}\sum_{u=1}^M \mathrm{diag}(\boldsymbol{s}_u) \boldsymbol{G} \boldsymbol{h}_{u,m}\!+ \! \bar{\boldsymbol{z}}_m  =\sqrt{p} \tilde{\boldsymbol{G}} \boldsymbol{h}_{m} \!+\!  \bar{\boldsymbol{z}}_m, ~ \forall m,
 \label{eqS3.1}
\end{align}
where $\boldsymbol{h}_{m}=[\boldsymbol{h}_{1,m},\ldots,\boldsymbol{h}_{M,m}]^T$, $\boldsymbol{G} \in \mathbb{C}^{N \times L} $ with the $(n,l)$-th element being $G_{n,l} = e^{\frac{-j2\pi (n-1)(l-1)}{N}}$, $\tilde{\boldsymbol{G}}=[\mathrm{diag}(\boldsymbol{s}_1) \boldsymbol{G},\ldots,$ $\mathrm{diag}(\boldsymbol{s}_M) \boldsymbol{G}]$, and $\bar{\boldsymbol{z}}_m= \boldsymbol{W} \boldsymbol{z}_m \sim \mathcal {CN}(0,\sigma_z^2 \boldsymbol{I})$.

In this paper, we assume that all the BSs know $\boldsymbol{s}_1,\ldots,\boldsymbol{s}_M$ sent by the BSs. For example, in the channel estimation phase for communication, $\boldsymbol{s}_m$'s are pilot signals and can be known by all the BSs. In the data transmission phase, the BSs can exchange the messages $\boldsymbol{s}_m$'s with each other over the fronthaul links as in cloud radio access network. Hence, $\tilde{\boldsymbol{G}}$ in (\ref{eqS3.1}) is known by all the BSs. Moreover, due to the limited number of targets, scatters, and clutters, very few elements in $\boldsymbol{h}_{m}$'s are non-zero, i.e., $\boldsymbol{h}_{m}$ is a sparse channel vector, $\forall m$. This motivates us to utilize the LASSO technique to estimate the time-domain channels by solving the following problem \cite{LASSO}
\begin{equation}
\begin{aligned}
 & \underset{{\boldsymbol{h}_{m}} }{\text{minimize}}
 ~ \frac{1}{2}\Vert \boldsymbol{\bar{y}}_{m} - \sqrt{p} \tilde{\boldsymbol{G}} \boldsymbol{h}_{m} \Vert_2^2+ \lambda  \Vert \boldsymbol{h}_{m} \Vert_1,
  \end{aligned}
 \label{eqS3.2}
\end{equation}
where $\lambda$ is a constant to control the sparsity of each $\boldsymbol{h}_{m}$. The above problem is convex and can be solved efficiently using CVX.

Let $\bar{\boldsymbol{h}}_{m}=[\bar{\boldsymbol{h}}_{1,m},\ldots,\bar{\boldsymbol{h}}_{M,m}]^T$ denote the optimal solution to problem (\ref{eqS3.2}), where $\bar{\boldsymbol{h}}_{u,m}=[\bar{h}_{u,m,1},\ldots,\bar{h}_{u,m,1}]^T$, $m,u=1,\ldots,M$. As discussed in Section \ref{sec:model}, if $\bar{h}_{u,m,l}\neq 0$ for some $l$, then we claim that there exists a path from BS $m$ to BS $u$ whose propagation delay is of $l$ OFDM sample periods. In this case, we estimate the range of this path as follows \cite{shi2022device}
\begin{equation}
 \bar{r}_{u,m,l}=\frac{(l-1)c_0}{N \Delta f} + \frac{c_0 }{2N \Delta f},~ {\rm if} ~ \bar{h}_{u,m,l}\neq 0,
 \label{eqS3.3}
\end{equation}where $\Delta f$ (in Hz) denotes the OFDM sub-carrier spacing such that $N \Delta f$ is the bandwidth. Based on the definitions of Type I, Type II, and Type III paths in Section \ref{sec:model}, if $\bar{h}_{u,m,l}\neq 0$, then $\bar{r}_{u,m,l}$ in (\ref{eqS3.3}) satisfies
\begin{equation}
 \bar{r}_{u,m,l}=
 \left \{
 \begin{array}{ll}
d_{u,m,k_{u,m,l}}+\epsilon_{u,m,k_{u,m,l}},&\text{Type I path}  \\
d_{u,m,k_{u,m,l}}+\epsilon_{u,m,k_{u,m,l}}+\eta_{u,m,l},&\text{Type II path},\\
\bar{\gamma}_{u,m,l},&\text{Type III path},
\end{array}
 \right.
\label{eqS3.4}
\end{equation}
where $k_{u,m,l}$ denotes the index of the target which reflects the signal from BS $u$ to BS $m$ with a delay of $l$ OFDM sample periods, $\epsilon_{u,m,k_{u,m,l}}$ denotes the error caused by the estimation shown in (\ref{eqS3.3}), $\eta_{u,m,l}$ denotes the bias introduce by the NLOS propagation, and $\bar{\gamma}_{u,m,l}$ denotes the estimated range of a Type III path with some clutter.

To summarize, after Phase I of our considered two-phase networked device-free sensing protocol, each BS $m$ will possess $M$ range estimation sets
\begin{align}
\mathcal{D}_{u,m}=\{\bar{r}_{u,m,l}|\forall l ~ {\rm with} ~ \bar{h}_{u,m,l} \neq 0\}, ~ u=1,\cdots, M.
 \label{eqS3.7}
\end{align}Then, each BS $m$ will transmit the above $M$ range sets to the central processor via the fronthaul links. Note that $\mathcal{D}_{u,m}$ contains the range information for all the paths (Types I, II, and III) from BS $u$ to BS $m$. However, for each element in $\mathcal{D}_{u,m}$, we do not know whether it is the range of a Type I path, a Type II path, or a Type III path. Moreover, even if $\bar{r}_{u,m,l}$ is identified to be the range estimation associated with a Type I path from BS $u$ to BS $m$, we do not know whether $\bar{r}_{u,m,l}$ is an estimation of $d_{u,m,1}$, $\ldots$, or $d_{u,m,K}$. Therefore, data association, NLOS mitigation, and clutter suppression should be jointly conducted by the central processor to localize the targets in Phase II of our considered protocol.

\section{Phase II: Joint Data Association, NLOS Mitigation, and Clutter Suppression}\label{sec:Phase II}
With the knowledge about $\mathcal{D}_{u,m}$'s, $\forall u,m$, the objectives of the central processor in Phase II are two-fold. First, it needs to estimate the number of targets in the network, i.e., $K$. Second, it needs to estimate the coordinates of the $K$ targets, i.e., $(x_k,y_k)$, $k=1,...,K$. For convenience, given any set $\mathcal{D}$, let $\mathcal{D}(g)$ denote its $g$-th largest element. Then, define $g_{u,m,k}$ as an integer such that $\mathcal{D}_{u,m}(g_{u,m,k})$ is the range estimation of the LOS path from BS $u$ to target $k$ and then to BS $m$, $\forall u,m,k$. Moreover, define $\mathcal{G}_k=\{g_{u,m,k}, \forall u,m\}$ as the solution of data association, NLOS mitigation, and clutter suppression for target $k$ to all the $M$ BSs, $k=1,\ldots,K$. If $\mathcal{G}_1,\ldots,\mathcal{G}_K$ can be found out, then the number of the targets can be known as well by checking how many LOS estimations are matched to the targets. Further, given $\mathcal{G}_k$, the location of each target $k$ can be estimated based on its range information $\mathcal{D}_{u,m}(g_{u,m,k})$'s, $\forall u,m$. In the following, we show how to estimate the number of the targets and their locations via a proper design of $\mathcal{G}_1,\ldots,\mathcal{G}_K$.

First, we define the conditions that a feasible solution of $\mathcal{G}_1,\ldots,\mathcal{G}_K$ should satisfy. Note that given $u,m$, the number of elements in the range set $\mathcal{D}_{u,m}$ is denoted by its cardinality $|\mathcal{D}_{u,m}|$. Therefore, the elements in $\mathcal{G}_1,\ldots,\mathcal{G}_K$ should satisfy
\begin{align}
g_{u,m,k}\in \{1,2,\ldots,|\mathcal{D}_{u,m}|\}, ~ \forall u,m,k.
\label{eqS4.1}
\end{align}Moreover, if one range estimation in $\mathcal{D}_{u,m}$ is matched to target $k$, then it cannot be matched to another user $\bar{k}\neq k$, i.e.,
\begin{align}
g_{u,m,k}\neq g_{u,m,\bar{k}}, ~ {\rm if} ~ \bar{k}\neq k, ~ \forall u,m.
\label{eqS4.2}
\end{align}Another condition that $\mathcal{G}_1,\ldots,\mathcal{G}_K$ should satisfy arises from (\ref{eqS2.2}): the length of the LOS path from BS $u$ to target $k$ to BS $m$, i.e., $d_{u,m,k}$, is equal to the sum of the distance between BS $u$ and target $k$, i.e., $d_{u,k}$, and that between BS $m$ and target $k$, i.e., $d_{m,k}$. Note that the imperfect estimations of $d_{u,m,k}$, $d_{u,k}$, and $d_{m,k}$'s are $\mathcal{D}_{u,m}(g_{u,m,k})$ (also $\mathcal{D}_{m,u}(g_{m,u,k})$), $\mathcal{D}_{u,u}(g_{u,u,k})/2$, and $\mathcal{D}_{m,m}(g_{m,m,k})/2$. Therefore, we set the following constraints for $\mathcal{G}_1,\ldots,\mathcal{G}_K$:
\begin{align}
& \left|\frac{\mathcal{D}_{u,u}(g_{u,u,k})}{2}\!+\!\frac{\mathcal{D}_{m,m}(g_{m,m,k})}{2}\!-\!\mathcal{D}_{u,m}(g_{u,m,k})\right| \leq \delta,
\label{eqS4.3} \\
& \left|\frac{\mathcal{D}_{u,u}(g_{u,u,k})}{2}\!+\!\frac{\mathcal{D}_{m,m}(g_{m,m,k})}{2}\!-\! \mathcal{D}_{m,u}(g_{m,u,k})\right| \leq \delta, ~ \forall m,u,k,
\label{eqS4.4}
\end{align}where $\delta>0$ is a given threshold. The last constraint about $\mathcal{G}_1,\ldots,\mathcal{G}_K$ is on the localization residual associated with this solution about data association, NLOS mitigation, and clutter suppression. Specifically, given $\mathcal{G}_k$, the location of target $k$ is estimated by solving the following nonlinear least squared (NLS) problem
\begin{align*}
{\rm (P1)} ~ \mathop{\textup{minimize}}_{x_k,y_k}~ \sum\limits_{u=1}^M\sum\limits_{m=1}^M &(f(x_k,y_k,a_u,b_u)+f(x_k,y_k,a_m,b_m)\nonumber \\ & -\mathcal{D}_{u,m}(g_{u,m,k}))^2,
\end{align*}where $f(x_k,y_k,a_m,b_m)$'s are given in (\ref{eqS2.1}). Problem (P1) is a non-convex problem. We can adopt the Gauss-Newton method to solve it \cite{Torrieri84}. Given $\mathcal{G}_k$ for target $k$, define $R(\mathcal{G}_k)$ as the value of problem (P1) achieved by the Gauss-Newton method. Therefore, $R(\mathcal{G}_k)$ can be interpreted as the residual for localizing target $k$ given $\mathcal{G}_k$. If $\mathcal{G}_k$ is the right solution, then the localization residual $R(\mathcal{G}_k)$ should be small, $\forall k$. We thus set the following residual constraints about $\mathcal{G}_k$'s:
\begin{align}
R(\mathcal{G}_k)\leq \beta, ~ k=1,\ldots,K, \label{eqS4.6}
\end{align}where $\beta>0$ is some given threshold.

To summarize, any $\mathcal{G}_1,\ldots,\mathcal{G}_K$ satisfying constraints (\ref{eqS4.1})-(\ref{eqS4.6}) can be a feasible solution for data association, NLOS mitigation, and clutter suppression. Note that if $\mathcal{G}_1,\ldots,\mathcal{G}_K$ is a feasible solution that satisfies constraints (\ref{eqS4.1})-(\ref{eqS4.6}), then $\mathcal{G}_2, \ldots, \mathcal{G}_K$ is also a new feasible solution, with target $1$ not detected. In this paper, we want to maximize the number of targets whose locations can be estimated. Therefore, the solution of data association solution, NLOS mitigation, and clutter suppression can be found by solving the following problem
\begin{align*}
{\rm (P2)} ~ \mathop{\textup{maximize}}_{K,\mathcal{G}_1,\ldots,\mathcal{G}_K}  & ~ K \\ \mathop{\textup{subject to}} ~ & ~  (\ref{eqS4.1})-(\ref{eqS4.6}).
\end{align*}

The above problem can be solved by exhaustive search, i.e., given each $K$ and $\mathcal{G}_1,\ldots,\mathcal{G}_K$, we check whether conditions (\ref{eqS4.1})-(\ref{eqS4.6}) hold. However, such an approach needs to solve the non-convex problem (P1) many times, which is of high complexity. To resolve this issue, we first ignore constraints (\ref{eqS4.2}) and (\ref{eqS4.6}) in problem (P2), which leads to the following problem
\begin{align*}
{\rm (P3)} ~ \mathop{\textup{maximize}}_{K,\mathcal{G}_1,\ldots,\mathcal{G}_K}  & ~ K \\ \mathop{\textup{subject to}} ~ & ~  (\ref{eqS4.1}), ~ (\ref{eqS4.3}), ~ (\ref{eqS4.4}).
\end{align*}Let $\bar{K}$ and $\bar{\mathcal{G}}_1,\ldots,\bar{\mathcal{G}}_{\bar{K}}$ denote the optimal solution to the above problem. Due to the sum distance constraints (\ref{eqS4.3}) and (\ref{eqS4.4}), the value of $\bar{K}$ is generally small, because the probability that (\ref{eqS4.3}) and (\ref{eqS4.4}) hold for Type II paths and Type III paths is very small. For convenience, define $\bar{\mathcal{G}}=\bar{\mathcal{G}}_1\cup\bar{\mathcal{G}}_2\cup \ldots\cup \bar{\mathcal{G}}_{\bar{K}}$. Then, we just need to check which subsets in $\bar{\mathcal{G}}$ satisfy constraint (\ref{eqS4.6}), i.e., the number of times to solve problem (P1) is significantly reduced compared to the exhaustive search approach to problem (P2). Define
\begin{align}
\tilde{\mathcal{G}}=\{\tilde{\mathcal{G}}_k|\tilde{\mathcal{G}}_k\in \bar{\mathcal{G}} ~ {\rm and} ~ R(\tilde{\mathcal{G}}_k)\leq \beta\}.
\label{eqS4.7}
\end{align}In other words, by removing $\bar{\mathcal{G}}_k$'s that do not satisfy condition (\ref{eqS4.6}) from $\bar{\mathcal{G}}$, we can obtain $\tilde{\mathcal{G}}$. Note that each subset contained in $\tilde{\mathcal{G}}$ satisfies conditions (\ref{eqS4.1}), (\ref{eqS4.3})-(\ref{eqS4.6}), i.e., it is a feasible solution of data association, NLOS mitigation, and clutter suppression to localize one target. However, for two subsets  $\tilde{\mathcal{G}}_k\in \tilde{\mathcal{G}}$ and $\tilde{\mathcal{G}}_{\bar{k}}\in \tilde{\mathcal{G}}$, it is possible that $g_{u,m,k}=g_{u,m,\bar{k}}$ for some $u,m$, i.e., in these two solutions, some estiamted range at BS $m$ is matched to both target $k$ and target $\bar{k}$. Therefore, the last step to solve problem (P2) is to select the maximum number of subsets in $\tilde{\mathcal{G}}$ such that condition (\ref{eqS4.2}) can be satisfied. Such a problem can be formulated as
\begin{align*}
{\rm (P4)} ~ \mathop{\textup{maximize}}_{K,\mathcal{G}_1,\ldots,\mathcal{G}_K}  & ~ K \\ \mathop{\textup{subject to}} ~ & ~  \mathcal{G}_k\in \tilde{\mathcal{G}}, ~ k=1,\ldots,K, \\ & ~ (\ref{eqS4.2}).
\end{align*}Because the number of subsets in $\tilde{\mathcal{G}}$ that satisfy so many conditions is small, problem (P4) can be efficiently solved. After this problem is solved, the number of targets is known, and we can solve problem (P1) to localize these $K$ targets.

\begin{remark}
It is worth noting that a similar two-phase networked device-free sensing strategy was considered in our recent work \cite{shi2022device}, where Type II and Type III paths are assumed to be absent such that NLOS mitigation and clutter suppression are not needed. Moreover, the sum distance constraints (\ref{eqS4.3}) and (\ref{eqS4.4}) are not utilized in the data association algorithm design. In this work, we point out that the sum distance constraints (\ref{eqS4.3}) and (\ref{eqS4.4}) can greatly reduce the number of feasible solutions such that data association, NLOS mitigation, and clutter suppression can be jointly designed with low complexity.
\end{remark}

\section{Numerical Results}
In this section, we provide numerical examples to verify the effectiveness of our considered two-phase networked device-free sensing scheme in the multi-path propagation environment. Specifically, the channel bandwidth is assumed to be $B=N\Delta f=400$ MHz. Moreover, we assume that $M=4$ BSs and $K\in [2,6]$ targets are uniformly distributed  in a 120 m $\times$ 120 m square, and randomly generate $10^4$ realizations of their locations. In each realization, we also randomly generate some Type II paths, i.e., the NLOS paths, and some Type III paths, i.e., the paths arising from clutters. Then, given the Types I, II, and III paths in each realization, we apply the two-phase protocol described in Sections \ref{sec:Phase I} and \ref{sec:Phase II} to estimate the number and the locations of the targets. Under our strategy, if target $k$ is not detected, or if target $k$ is detected, but the estimated location of it is not lying within a radius of $r=0.375$ m from its true location, then we claim that a missed detection event occurs for target $k$. On the other hand, if a target that does not exit is detected, we claim that a false alarm event occurs. At the $i$-th iteration, let $N_i$ and $T_i$ denote the numbers of the missed detection events and the false alarm events. Then, over the $10^4$ realizations, the probabilities of missed detection and false alarm are defined as $P_{{\rm MD}}=\sum_{i=1}^{10^4}N_i/(K\times 10^4)$ and $P_{{\rm FA}}=\sum_{i=1}^{10^4}T_i/(K\times 10^4)$.

\begin{figure}
\centering
\includegraphics[scale=0.6]{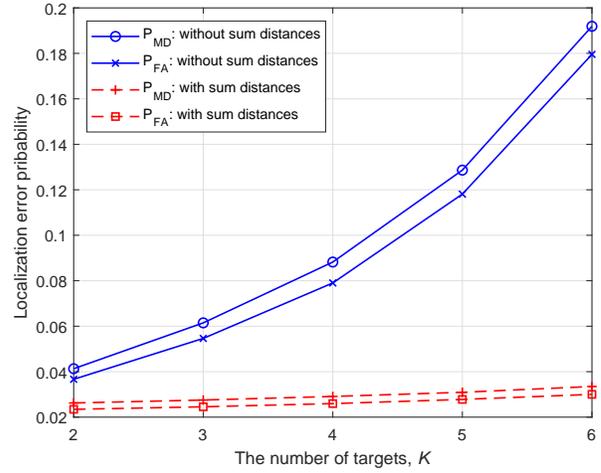}
\caption{The missed detection and false alarm performance.}
\label{fig2}\vspace{-10pt}
\end{figure}

Fig. \ref{fig2} shows the performance of our proposed two-phase networked device-free sensing scheme, when $K$ ranges from $2$ to $6$. For the performance benchmark, we select the scheme in \cite{shi2022device}, where the sum distances, i.e., $d_{u,m,k}$'s, are not utilized for data association. It is observed from Fig. \ref{fig2} that under our proposed scheme, the probabilities of missed detection and false alarm are below $3.4\%$ when $K$ ranges from $2$ to $6$. Moreover, it is also observed that the utilization of the sum distances for joint data association, NLOS mitigation, and clutter suppression can significantly reduce the probabilities of missed detection and false alarm compared to the scheme proposed in \cite{shi2022device}. At last, because the constraints associated with the sum distances, i.e., (\ref{eqS4.3}) and (\ref{eqS4.4}), can significantly reduce the number the times to solve NLS problem (P1), which is non-convex, the strategy proposed in this paper is observed to generate the localization solution with a significantly reduced CPU running time compared to the strategy proposed in \cite{shi2022device}.

\section{Conlusion}
In this paper, we considered the networked device-free sensing scheme in the 6G network under a multi-path propagation environment. A two-phase localization protocol was studied. In the first phase, range estimation was performance with the OFDM channel estimation techniques. In the second phase, we proposed an efficient algorithm for joint data association, NLOS mitigation, and clutter suppression. Then, the number and the locations of the targets are estimated based on the range estimations from the LOS paths. Numerical results showed that our proposed strategy can enable the BSs to accurately localize the targets with small probabilities of missed detection and false alarm.

\bibliographystyle{IEEEbib}
\bibliography{ref}

\end{document}